\documentclass{optica-article}

\journal{opticajournal} 

\articletype{Research Article}

\begin{document}

\title{Multi-mode Photonic Time Crystals Based on Time-Modulated Metasurface Waveguides}

\author{Z.~Li,\authormark{1} M.~S.~Mirmoosa,\authormark{2} V.~Asadchy,\authormark{3} and X. Wang\authormark{4,*}}

\address{\authormark{1}College of Information and Communication Engineering, Harbin Engineering University, China\\
\authormark{2}Department of Physics and Mathematics, University of Eastern Finland, P.O.~Box~111, FI-80101 Joensuu, Finland\\
\authormark{3}Department of Electronics and Nanoengineering, Aalto University, P.O.~Box~15500, FI-00076~Aalto, Finland\\
\authormark{4}College of Physics and Optoelectronic Engineering, Harbin Engineering University, China}

\email{\authormark{*}xuchen.wang@hrbeu.edu.cn} 


\begin{abstract*} 
Photonic time crystals are electromagnetic media with periodically time-varying parameters, enabling momentum band gaps, parametric amplification, and frequency conversion beyond what is possible in time-invariant systems. 
So far, they have been explored mainly in single-mode systems, which limits the range of accessible physical phenomena. 
Here, we introduce an impenetrable metasurface waveguide as a multimode time-varying platform supporting both guided surface modes and higher-order guided volume modes. 
We show that temporal modulation in this platform gives rise not only to conventional intramodal band gaps associated with same-branch coupling, but also to tilted intermodal band gaps originating from coupling between different guided-mode branches. 
Unlike intramodal band gaps, these intermodal band gaps are not restricted to half the modulation frequency and can enable directional wave amplification, where the amplified field carries energy along the waveguide even inside the band gap. 
We further show that the modulation phase difference provides an effective symmetry-control parameter: by exploiting temporal glide symmetry, one can selectively suppress or enhance gap opening for interactions between modes of the same or different symmetry. 
These results establish a versatile multimode platform for photonic time crystals, offering one of the simplest and most experimentally accessible routes to tilted band gaps compared with volumetric dispersive PTC implementations and, more broadly, opening new opportunities for time-varying electromagnetic systems.

\end{abstract*}

\section{INTRODUCTION}
To enhance flexibility in controlling electromagnetic waves, significant research has been conducted on time-varying systems in recent years~\cite{engheta2020metamaterials, hayran2023using}. These studies have introduced novel concepts, such as time reflection and time refraction, which arise from abrupt changes in constitutive parameters (relative to the oscillation period), including permittivity. In addition, several intriguing phenomena induced by temporal discontinuities have been reported, including temporal aiming~\cite{pacheco2020temporal}, antireflection temporal coatings~\cite{pacheco2020antireflection}, and energy canalization~\cite{ptitcyn2025temporal}. These advancements underscore the potential of time as a new dimension for electromagnetic wave control. Beyond temporal discontinuity, another form of time variation that has garnered more attention is periodic time modulation, where certain parameters of the system change periodically over time. For example, the permittivity and (or) permeability of materials are modulated by a periodic square function~\cite{martinez2016temporal, wang2018photonic}. This periodic time-varying system is referred to as a photonic time crystal (PTC), which exhibits momentum band gaps analogous to the energy band gaps of spatial photonic crystals~\cite{asgari2024theory}. These momentum band gaps have been shown to amplify incident waves~\cite{lyubarov2022amplified}, which is the most distinctive feature of photonic time crystals. Besides this hallmark, diverse physical phenomena have been reported across different types of PTC platforms, including temporal Smith-Purcell effect~\cite{zhu2025smith}, subluminal Cherenkov radiation~\cite{dikopoltsev2022light}, broad band gaps enabled by ghost waves~\cite{dong2025nonuniform}, and super absorber that surpass the Rozanov bound~\cite{ciabattoni2025observation}.

For now, most studies on photonic time crystals have been limited to non-dispersive single-mode bulk media, which require ultra-fast material modulations. Recent works \cite{feng2024temporal} and~\cite{ozlu2025floquet} demonstrated that periodic time modulation in Lorentz-type dispersive bulk media, could lead to the emergence of intermodal band gaps that do not require very high modulation frequencies (usually twice the frequency of the incident wave) due to the interaction between two different bands in Lorentz media. Nevertheless, implementing time modulation in bulk media remains challenging due to the complexity of the pumping schemes, which require a sufficiently large volumetric modulation region within the material.
To simplify the modulation scheme and probing of the signal, photonic time crystals realized on a metasurface platform were proposed~\cite{wang2023metasurface,wang2025expanding,garg2025photonic}. Metasurfaces are a two-dimensional form of metamaterials composed of subwavelength unit cells. It was demonstrated that metasurface-based PTCs, despite their simpler topology, retain the essential physics of volumetric photonic time crystals and support a shared momentum band gap for both surface and free-space electromagnetic waves~\cite{wang2023metasurface}.
However, metasurface-based PTCs reported to date support only band gaps arising from the crossing of two identical modes, thereby limiting the available degrees of freedom for exploring the complex physics of time-varying materials and requiring a high frequency of the temporal modulations.

In this paper, we introduce a multimode PTC based on a time-varying impenetrable metasurface waveguide to overcome the limitations of existing platforms with restricted inter-modal interactions. This parallel-plate metasurface waveguide configuration can support not only guided surface modes but also multiple higher-order guided volume modes, thereby enabling richer inter-modal interactions. This gives rise to wave phenomena inaccessible in conventional volumetric PTCs and metasurface-based PTCs. 
We use the plane-wave expansion method to derive the band structure of the time-varying metasurface waveguides. Theoretical derivations reveal that under time modulation, coupling occurs between different modes of the PTC. Band crossings formed by the same single mode result in intramodal band gaps located at half the modulation frequency, while the band crossings formed by two different modes lead to intermodal band gaps. These band gaps do not necessarily occur at half the modulation frequency and are tilted in the frequency-momentum plane, exhibiting exotic directional wave parametric amplification. Furthermore, the opening and closing of these band gaps can be flexibly controlled by tuning the phase difference between the temporal modulations of the two impenetrable metasurfaces. These results establish a versatile multimode platform for PTCs, offering one of the simplest and most experimentally accessible reported routes to tilted band gaps compared with volumetric dispersive PTC implementations, and extend time-varying electromagnetics beyond the single- and dual-mode regimes.

\section{DERIVATION AND DISCUSSION OF DISPERSION RELATIONS IN TIME-INVARIANT CASE}  

\begin{figure}[h]  
    \centering 
    \includegraphics[width=1\textwidth]{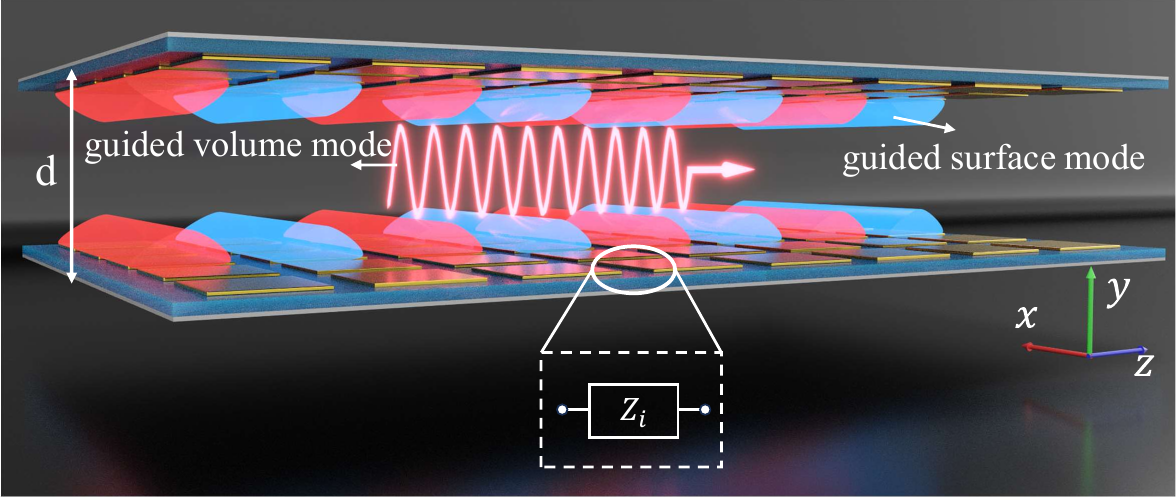} 
    \caption{Geometry of the waveguide formed by two impenetrable metasurfaces. Each meta-atom can be described generally by a time-varying impedance $Z_i$, where $i=1,2$. In the following discussion, the impedance of the upper sheet and lower sheet is denoted by $Z_1$ and $Z_2$, respectively.} 
    \label{fig:1} 
\end{figure}

Before analyzing the time-varying case, we first consider the time-invariant case. The proposed waveguide structure under study is shown in Fig.~\ref{fig:1}. Two impenetrable metasurfaces are placed facing each other at $y=0$ and $y=d$, with free space between them. An important assumption, which is also the basis of this study, is that the metasurfaces are isotropic and reactive, meaning that we can model each metasurface as an equivalent reactive sheet, for example, a capacitive or inductive sheet. For the convenience of formula derivation, it is necessary to first define the propagation direction. Throughout this paper, the $+z$ direction is assumed to be the direction of wave propagation, while the $xy$-plane is the transverse plane. The metasurfaces are uniform in the $x$-direction.

We first assume that both metasurface sheets are capacitive and time-invariant, with the capacitance of the upper sheet being $C_1$ and the lower sheet being $C_2$. According to the study in \cite{tretyakov2003analytical, ma2019parallel}, two capacitive metasurfaces only support TE modes. Therefore, we assume the modes are TE-polarized. It is worth noting, however, that two inductive metasurfaces can support TM-polarized guided surface modes.

The supported modes of the proposed waveguide are found as source-free eigenmodes satisfying the waveguide boundary conditions. For TE modes in a parallel-plate waveguide, the longitudinal magnetic-field component satisfies the scalar Helmholtz equation~\cite{pozar2021microwave}. Similar to the derivation of the electromagnetic field in a metallic parallel-plate waveguide, we can express the tangential component of the magnetic field $H_z$ inside the structure as:
\begin{equation}
    H_{z,\mathrm{TIV}}=(H_{+} e^{-j k_c y}+H_{-} e^{j k_c y}) e^{-j \beta z} e^{j \omega t}. \label{eq.1}
\end{equation}
Here, $k_c=\sqrt{k_0^2-\beta^2}$ where $k_0$ represents the free-space wave number, and $\beta$ represents the propagation constant along $z$ direction.
Then substituting $H_z$ into Maxwell's curl equation ($\nabla \times \mathbf{E}=-\frac{\partial \mathbf{B}}{\partial t}$) in the absence of sources, we obtain the tangential component of electric field $E_x$:

\begin{equation}
    E_{x,\mathrm{TIV}}=\frac{\omega \mu_0}{k_c} \left(-H_{+} e^{-j k_c y}+H_{-} e^{j k_c y}\right) e^{-j \beta z}e^{j \omega t} \label{eq.2}
\end{equation}

\begin{figure}[h]
    \centering
    \includegraphics[width=1\linewidth]{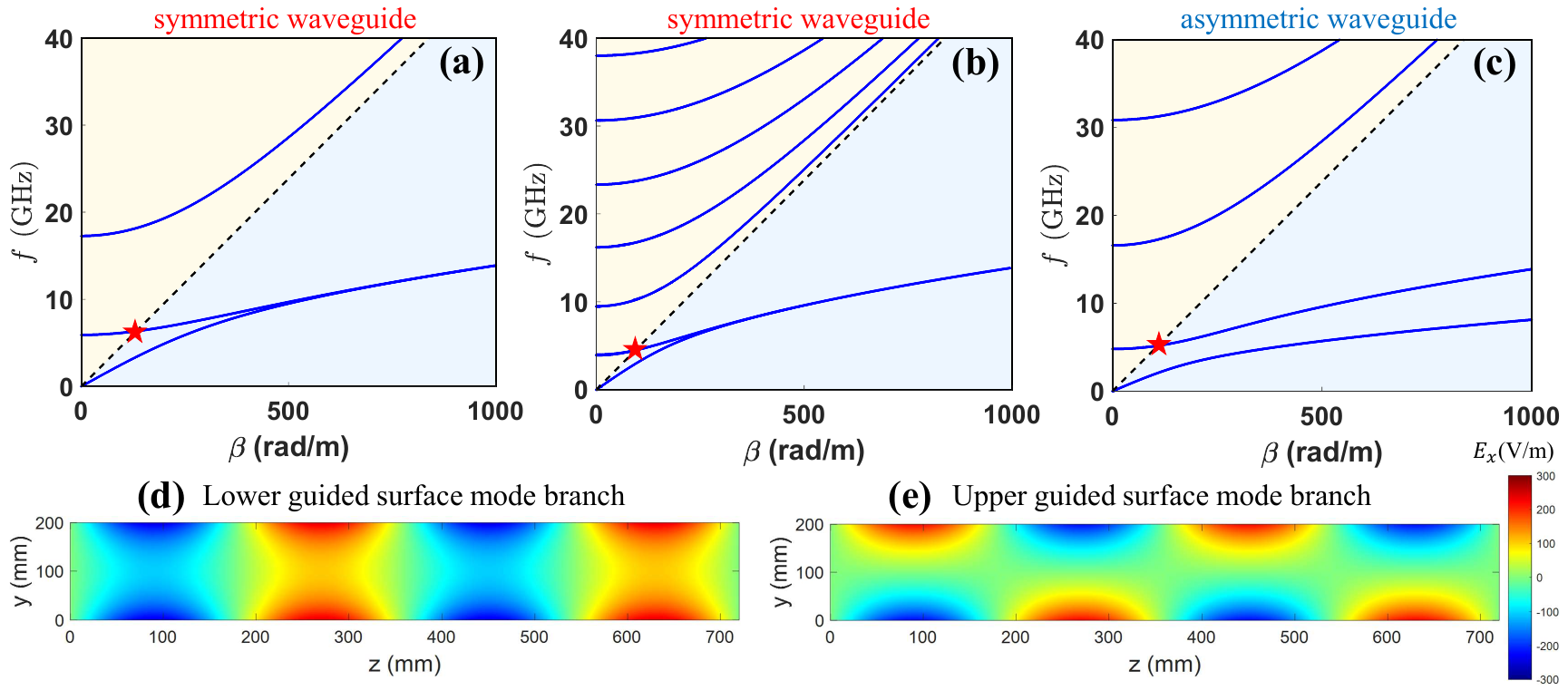}
    \caption{(a)--(c) Dispersion curves of the waveguide formed by two impenetrable time-invariant metasurfaces with surface capacitances $C_1$ and $C_2$. (a) $C_1=C_2=0.1~\mathrm{pF}$, $d=10~\mathrm{mm}$. (b) $C_1=C_2=0.1~\mathrm{pF}$, $d=20~\mathrm{mm}$. (c) $C_1=0.1~\mathrm{pF}$, $C_2=0.3~\mathrm{pF}$, $d=10~\mathrm{mm}$. The black dashed line represents the light line. The ``cutoff'' frequencies of the upper surface modes are marked by red stars. (d)--(e) Electric-field profiles of the lower and upper guided surface modes, respectively. }
    \label{fig:2}
\end{figure}

To find the unknowns $H_{-}$ and $H_{+}$ in Eq.~(\ref{eq.1}) and Eq.~(\ref{eq.2}), we need to apply the impedance boundary conditions on the two reactive boundaries:
\begin{subequations}
\begin{align}
    Z_1(-\Vec{e_y}\times H_z \Vec{e_z})=E_x \Vec{e_x}\\
   Z_2(\Vec{e_y}\times H_z\Vec{e_z})=E_x \Vec{e_x}, \label{eq.3}
\end{align} \label{eq.32}
\end{subequations}
where $Z_1=1/j\omega C_1$, $Z_2=1/j\omega C_2$. Substituting Eqs.~(\ref{eq.1}) and (\ref{eq.2}) into (\ref{eq.32}), we finally obtain:

\begin{equation}
    \begin{bmatrix}
    &\left(-1+\frac{\omega \mu_0}{Z_1 k_c}\right) e^{-j k_c d} &-\left(1+\frac{\omega \mu_0}{Z_1 k_c}\right) e^{j k_c d}\\
    &1+\frac{\omega \mu_0}{Z_2 k_c} &1-\frac{\omega \mu_0}{Z_2 k_c}
\end{bmatrix}
\left[
\begin{array}{c}
    H_+\\
    H_-
\end{array}
\right]=0. \label{eq.4}
\end{equation}
Equation~(\ref{eq.4}) shows that 
$\overline{\overline{A}}\cdot\vec{H}=\vec{0}$, where 
$\overline{\overline{A}}$ is a $2\times2$ matrix depending on $\omega$ and $\beta$, and 
$\vec{H}$ is the column vector of the complex magnetic-field amplitudes. 
A nonzero field solution exists only when
$\det\overline{\overline{A}}=0$, which gives the dispersion relation of the waveguide.
The resulting dispersion curves are shown in Fig.~\ref{fig:2}. 
Two qualitatively different types of guided modes are supported. 
The branches below the light line (dashed black line in Fig.~\ref{fig:2}(a)), $\beta>k_0=\omega/c$, correspond to \textit{guided surface} modes. 
For these modes, the transverse wavenumber is imaginary, so the fields are evanescent in the transverse direction and are localized near the reactive metasurface boundaries (decay exponentially toward the waveguide center). 
By contrast, the branches above the light line, $\beta<k_0$, correspond to \textit{guided volume} (parallel-plate-type) modes. 
For these modes, the transverse wavenumber is real, so the fields oscillate across the spacing between the two impenetrable metasurfaces, similarly to higher-order modes of a parallel-plate waveguide. 
Importantly, these volume modes are still guided modes of the present structure, because the impenetrable metasurface boundaries confine the fields inside the waveguide.
This modal structure differs from previous metasurface platforms for PTC realizations~\cite{wang2023metasurface,lin2024temporally,shahriar2024idea,sisler2024electrically,moreno2024space}, where the relevant guided spectrum was limited to surface-type modes. 
In this sense, the considered metasurface waveguide is a richer multimode platform. 
Temporal modulation can induce both \textit{intramodal} coupling, where different Floquet harmonics of the same guided-mode branch interact, and \textit{intermodal} coupling, where Floquet harmonics belonging to different branches interact. 
In particular, the latter enables coupling between guided surface modes and guided volume modes, giving rise to mode interactions that are not available in single-branch or surface-mode-only platforms.

As is seen from Fig.~\ref{fig:2}(a), the upper guided surface branch suffers from a ``cutoff'' frequency which can be readily obtained by letting $k_c$ in Eq.~(\ref{eq.4}) approach zero, namely, $f_{\mathrm{cutoff}} \approx \frac{1}{2 \pi} \sqrt{\frac{C_1+C_2}{\mu_0 d C_1 C_2}}$. When the frequency is higher than this ``cutoff'' frequency, it corresponds to a guided surface mode; otherwise, it becomes a guided volume mode. Therefore, this ``cutoff'' frequency is in fact the transition point between the two types of  modes. In addition to the ``cutoff'' frequency, the distance between the two sheets and the sheet impedances of the two metasurfaces also play  important roles in shaping the dispersion curve. As can be seen from Fig.~\ref{fig:2}(b), when the distance $d$ increases,  more guided volume modes appear. 

Additionally, the mode patterns of the two guided surface modes differ: the upper one is an even (symmetric with respect to the $y$-axis) mode, while the other one is an odd mode (antisymmetric with respect to the $y$-axis), see Fig.~\ref{fig:2}(d) and~\ref{fig:2}(e). When the two sheet impedances are identical, the dispersion curve shows that the two guided surface modes seem to gradually merge. However, both modes still exist; it is just that the system becomes symmetric and the even/odd modes degenerate, so their dispersion curves nearly coincide. By contrast, Fig.~\ref{fig:2}(c) shows that when the two sheet impedances are not identical, the two guided surface modes will split. 

\section{DERIVATION AND ANALYSIS OF BAND STRUCTURE IN TIME-VARYING CASE} 

\subsection{Derivation of Band Structure}

Based on the above multi-band dispersion curves of a stationary waveguide system, which provides rich opportunities for mode interactions, we now study the band structure when the effective reactances of the two sheets are periodically modulated in time. We assume that two capacitive sheets are periodically modulated in time, i.e., $C(t)=C(t+T_\mathrm{m})$, where $T_\mathrm{m}$ is the modulation period. For such a periodically time-varying system, plane-wave expansion method based on Floquet theory can be utilized to derive the band structure. First, it is straightforward to  expand $C(t)$ into a Fourier series as, $C(t)=\sum_p c_{p} e^{j p\omega_\mathrm{m} t}$ where $\omega_\mathrm{m} = 2\pi/T_\mathrm{m}$ is the modulation frequency and $c_{p}$ are the Fourier coefficients. Then, according to Floquet theory, periodic time modulation generates infinite numbers of frequency harmonics in the form of $\omega_n=\omega+n \omega_\mathrm{m}$, where $\omega$ is the Floquet frequency and $n$ is the Floquet order. It is also well known that, for such a spatially uniform  system that preserves translational symmetry along the $z$-direction,  the momentum  of the electromagnetic modes along the $z$-direction, i.e., propagation constant $\beta$, is conserved~\cite{asgari2024theory, galiffi2022photonics, ortega2023tutorial} under pure time modulation. It should be clarified that although the metasurfaces are spatially periodic along propagation direction, the spatial periodicity is sub-wavelength and consequently, can be treated as spatially uniform metasurfaces. Thus, all the Floquet harmonics share the same momentum $\beta$. Accordingly, the tangential components of the electric and magnetic fields of the eigenmodes can be expressed as the superposition of  infinitely many harmonics as shown below,
\begin{equation}
\begin{aligned}
    &H_{z,\mathrm{TV}}=\sum_n (H_{+,n} e^{-j k_{cn} y}+H_{-,n} e^{j k_{cn} y})e^{-j \beta z} e^{j \omega_n t}\\
    &E_{x,\mathrm{TV}}=\sum_n \frac{\omega_n \mu_0}{k_{cn}}(-H_{+,n} e^{-j k_{cn} y}+H_{-,n} e^{j k_{cn} y})e^{-j \beta z} e^{j \omega_n t}
\end{aligned} \label{eq.6}
\end{equation}

Then again using eigenmode analysis and applying time-varying boundary conditions, which is discussed in detail in \cite{wang2023metasurface} to each time modulated metasurface, we obtain:

\begin{equation}
\begin{aligned}
    &-\sum_n \frac{1}{j\omega_n}(H_{+,n} e^{-j k_{cn} d}+H_{-,n} e^{j k_{cn} d})e^{-j \beta z} e^{j \omega_n t} 
    =\\ &\sum_n \sum_p \frac{\omega_n \mu_0 c_{1p}}{k_{cn}}(-H_{+,n} e^{-j k_{cn} d}+H_{-,n} e^{j k_{cn} d})e^{-j \beta z} e^{j \omega_{n+p} t} 
\end{aligned} \label{eq.7}  
\end{equation}
for the upper metasurface located at $y=d$. And

\begin{equation}
\begin{aligned}
       \sum_n \frac{1}{j\omega_n}(H_{+,n}+H_{-,n})e^{-j \beta z} e^{j \omega_n t}=\sum_n \sum_p \frac{\omega_n \mu_0 c_{2p}}{k_{cn}}(-H_{+,n}+H_{-,n})e^{-j \beta z} e^{j \omega_{n+p} t}
\end{aligned} \label{eq.8}
\end{equation}

for the bottom metasurface located at $y=0$. Here, $c_{1p}$ and $c_{2p}$ represent the Fourier coefficients of temporal modulation of each metasurface. Then considering that the ranges of $n$ and $p$ are from $-\infty$ to $+\infty$, substituting $n-p$ for $n$ does not change the result of the summation. Thus, we rewrite the right-hand sides of Eqs.~(\ref{eq.7}) and (\ref{eq.8}) as: 

\begin{equation}
\begin{aligned}
    &-\sum_n \frac{1}{j\omega_n}(H_{+,n} e^{-j k_{cn} d}+H_{-,n} e^{j k_{cn} d})e^{j \omega_n t} 
    =\\ &\sum_n\sum_p \frac{\omega_{n-p} \mu_0 c_{1p}}{k_{c(n-p)}} (-H_{+,n-p} e^{-j k_{c(n-p)} d}+H_{-,n-p} e^{j k_{c(n-p)} d})e^{j \omega_n t}
\end{aligned} \label{eq.9}
\end{equation}

\begin{equation}
\begin{aligned}
    \sum_n \frac{1}{j\omega_n}(H_{+,n}+H_{-,n}) e^{j \omega_n t}=\sum_n \sum_p \frac{\omega_{n-p} \mu_0 c_{2p}}{k_{c(n-p)}}(-H_{+,n-p}+H_{-,n-p}) e^{j \omega_{n} t}
\end{aligned} \label{eq.10}
\end{equation}

Now, both sides of Eqs.~(\ref{eq.9}) and (\ref{eq.10}) share the same basic frequency, allowing us to obtain the relationship where the corresponding coefficients are equal. Thus, we have:

\begin{subequations}
\begin{align}
    -\frac{1}{j\omega_n \mu_0}(H_{+,n} e^{-j k_{cn} d} + H_{-,n} e^{j k_{cn} d}) &=  
    \sum_p \frac{\omega_{n-p} c_{1p}}{k_{c(n-p)}} (-H_{+,n-p} e^{-j k_{c(n-p)} d} + H_{-,n-p} e^{j k_{c(n-p)} d})\\
    H_{+,n} + H_{-,n} &= j\omega_n \mu_0 \sum_p \frac{\omega_{n-p} c_{2p}}{k_{c(n-p)}}(-H_{+,n-p} + H_{-,n-p})
\end{align} \label{eq.11}
\end{subequations}

After some mathematical transformations, we can rewrite Eqs.~(\ref{eq.11}a) and~(\ref{eq.11}b) into matrix form if we only consider harmonic order from $-N$ to $+N$, and finally obtain two equations regarding the column vector of unknown amplitudes $H_-$ and $H_+$ (the detailed matrix can be found in Appendix B):
\begin{subequations}
\begin{align}
   &(\Bar{\Bar{D}}-\Bar{\Bar{B}}\Bar{\Bar{W_1}})\Vec{H}_++(\Bar{\Bar{D}}^{-1}+\Bar{\Bar{B}}\Bar{\Bar{W_2}})\Vec{H}_-=0\\
    &(\Bar{\Bar{I}}+\Bar{\Bar{B}}\Bar{\Bar{A}})\Vec{H}_++(\Bar{\Bar{I}}-\Bar{\Bar{B}}\Bar{\Bar{A}})\Vec{H}_-=0
\end{align}
\end{subequations}

Here $\Bar{\Bar{I}}$ is a $(2N+1)\times(2N+1)$ identity matrix, $\Bar{\Bar{B}}$ and $\Bar{\Bar{D}}$ are $(2N+1)\times(2N+1)$ diagonal matrices, $\Bar{\Bar{A}}$, $\Bar{\Bar{W_1}}$ and $\Bar{\Bar{W_2}}$ are $(2N+1)\times(2N+1)$ Toeplitz matrices, and $\Vec{H}_+$ and $\Vec{H}_-$ are $(2N+1)\times 1$ column vector indicating the complex amplitudes of magnetic fields. Thus we finally derive the dispersion relation in time-varying case by setting the determinant of matrix $\Bar{\Bar{K}}$ to zero, where $\Bar{\Bar{K}}$ is a $(4N+2)\times(4N+2)$ block matrix: 
\begin{equation}
   \Bar{\Bar{K}}= \begin{bmatrix}
    \Bar{\Bar{D}}-\Bar{\Bar{B}}\Bar{\Bar{W_1}} &\Bar{\Bar{D}}^{-1}+\Bar{\Bar{B}}\Bar{\Bar{W_2}}\\
    \Bar{\Bar{I}}+\Bar{\Bar{B}}\Bar{\Bar{A}} &\Bar{\Bar{I}}-\Bar{\Bar{B}}\Bar{\Bar{A}}
\end{bmatrix}. \label{eq.15}
\end{equation}

This equation can be used to calculate the band structure of the metasurface waveguide PTC.

\subsection{Analysis of the Band Structure and Wave Behavior inside the Momentum Band Gaps}

This section demonstrates several notable  phenomena enabled by the time-varying metasurface waveguide, including directional amplification and controllable momentum band gaps. In such a multimode system, time modulation can produce momentum band gaps through different mode interaction processes. 

In what follows, we assume that the two metasurfaces are sinusoidally modulated in time according to
\begin{equation}
C_1(t)=C_{01}\bigl(1+m_1 \cos(\omega_\mathrm{m} t+\phi)\bigr)
\end{equation}
and
\begin{equation}
C_2(t)=C_{02}\bigl(1+m_2 \cos(\omega_\mathrm{m} t)\bigr),
\end{equation}
where \(\phi\) denotes the modulation phase difference between the two metasurfaces. 

\subsubsection{Intramodal (conventional) band gaps}

\begin{figure}[h]
    \centering
    \includegraphics[width=1\linewidth]{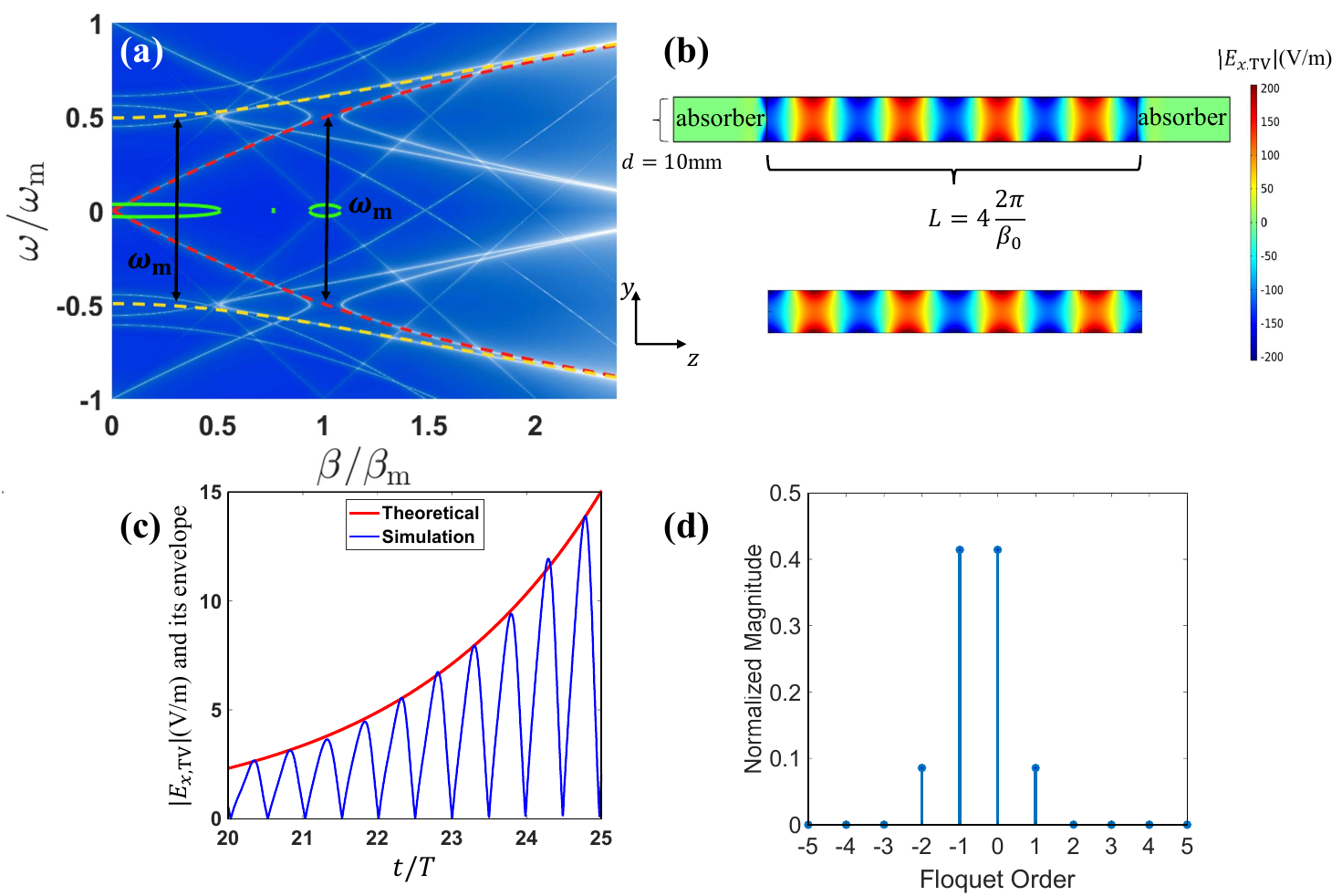}
    \caption{(a) Band structure of the PTC featuring intramodal band gaps. Green line represents imaginary part of eigenfrequency while bright lines represent real part. The dashed lines show the dispersion curves of the same metasurface waveguide but without temporal modulations. (b) Simulated (top) and theoretical  (bottom) result for the distribution of the transverse electric field $E_{x,\mathrm{TV}}$ in the momentum band gap inside the waveguide in the $yz$-plane. Here, $\beta_0=1.1\beta_{\mathrm{m}}$ and  $L$ represents the length of the metasurface waveguide, while $d$ represents the distance between the two metasurfaces. (c) Simulation result vs theoretical result for the temporal evolution of $E_{x,\mathrm{TV}}$. The blue line shows the simulated instantaneous magnitude, while the red line represents the analytical envelope; (d) Normalized magnitudes of different harmonics within the eigenmode at the center of the intramodal band gap  with $\beta_0=1.1\beta_{\mathrm{m}}$).}
    \label{fig:band strcuture}
\end{figure}

As a representative example, we set \(d=10~\mathrm{mm}\), \(C_{01}=C_{02}=0.1~\mathrm{pF}\), \(\omega_\mathrm{m}=2\pi\times12~\mathrm{GHz}\), \(\phi=0\), and \(m_1=m_2=0.3\). The calculated band structure is shown in Fig.~\ref{fig:band strcuture}(a).
As one can see, two intramodal band gaps appear around  the frequency of 6 GHz ($\omega_{\rm m}/2$). 
We refer to these gaps as intramodal
band gaps because each of them originates from the coupling of
Floquet harmonics associated with the same guided-mode branch.
This can be seen by folding the dispersion curves of the
time-invariant metasurface waveguide, shown by the dashed lines in
Fig.~\ref{fig:band strcuture}(a), into the first temporal Brillouin
zone of width $\omega_{\mathrm{m}}$. The crossings between the
folded copies of the same branch are then opened by the temporal
modulation, giving rise to the \textit{conventional} momentum band gaps.
More specifically, the left band gap is associated with intramodal coupling of the guided volume-mode branch with its negative-frequency counterpart, whereas the right band gap is associated with intramodal coupling of the lower guided surface-mode branch with its negative-frequency counterpart.

These band gaps exhibit the same essential features as those reported in non-dispersive time-modulated structures~\cite{lustig2023photonic,lustig2018topological}. However, their widths are significantly different with one being  much larger  than the other. 
To understand the origin of this band gap size contrast, we examine the static dispersion relation in Fig.~\ref{fig:2}(a). Around the cutoff frequency, the first guided volume mode branch is much flatter than the lower guided surface mode branch. This large band gap originates precisely from the folding of this upper branch. As discussed in Ref.~\cite{wang2025expanding}, the size of momentum band gap can be estimated from the static dispersion curve: a flatter branch generally favors the formation of a wider band gap.  Moreover, this larger momentum band gap  covers a certain range of $\beta$ starting from zero, indicating that a non-propagating mode can also be amplified. 

Next, we show how the wave is amplified inside the intramodal band gap. Numerical simulations were performed using the transient electromagnetic solver in COMSOL.  We choose the wavenumber at the center of the intramodal band gap, i.e. $\beta_0=1.1\beta_{\mathrm{m}}$. Since this band gap is formed through the interaction associated  with the lower branch of guided surface mode, we first excite the corresponding  guided surface mode before the modulation is introduced. As shown in Fig.~\ref{fig:band strcuture}(b), the simulated mode profile agrees very well with the theoretical prediction and, as expected, closely resembles the corresponding static-case mode shown in Fig.~\ref{fig:2}(d). At the moment of $t=20T$ where $T$ is the period of the incident guided surface wave,  the capacitance modulation is switched on while the external  excitation is simultaneously turned on. The modulation lasts for five cycles, during which the wave amplification is  observed.
Moreover, the growth rate of the  field amplitude magnitude agrees exactly with the theoretical prediction, exhibiting an exponential increase, as compared in Fig.~\ref{fig:band strcuture}(c). 

It is worth noting that an intramodal band gap is formed by the crossing of two Floquet copies of the same dispersion branch. Therefore, it occurs at half the modulation frequency, $\omega_{\mathrm{m}}/2$. Our calculations further show that, inside this gap, the spectral response is dominated by the harmonics at $\omega_{\mathrm{m}}/2$ and $-\omega_{\mathrm{m}}/2$, which have equal amplitudes, as illustrated in Fig.~\ref{fig:band strcuture}(d). Here, the magnitude of each harmonic is normalized by the transverse average of the electric-field profile inside the waveguide, $\frac{1}{d}\int_0^d E_{x,\mathrm{TV}}(y)\,dy$.
Since the two dominant harmonics have the same propagation constant $\beta$ but opposite frequencies, their phase velocities are opposite. Together with their equal amplitudes, this implies that the field inside the intramodal band gap forms a standing wave. 
As a result, the cycle-averaged net energy flux along the propagation direction vanishes, while the group velocity is not an appropriate measure of energy transport in PTCs~\cite{lee2026energy}.
Therefore, although the field amplitude can grow in time due to temporal modulation, the wave does not carry net energy along the propagation direction.



\begin{figure}[h] 
    \centering
    \includegraphics[width=0.9\linewidth]{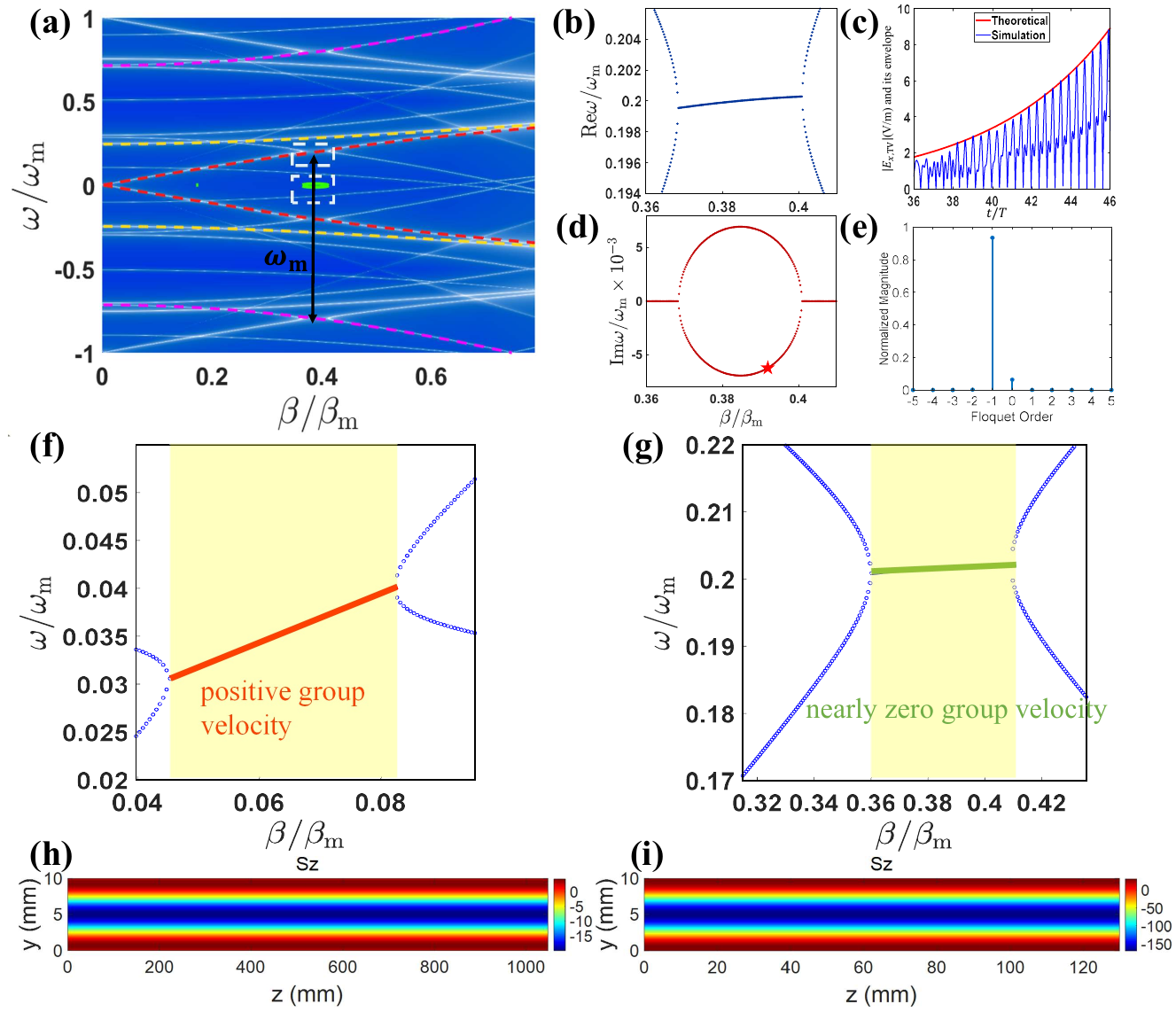}
    \caption{Band structure and wave behavior within the intermodal band gaps. (a) Band structure of the PTC featuring intermodal band gaps. Green lines represent the imaginary part of the eigenfrequency, while bright lines represent the real part. The dashed lines show the folded dispersion curves of the same metasurface waveguide without temporal modulations. (b) and (d) Zoomed-in views of the real and imaginary parts of the eigenfrequency within an intermodal band gap, respectively. The red star marks the specific eigenfrequency selected for numerical simulation. (c) Comparison between the simulated and theoretical results for the temporal evolution of the transverse electric field $E_{x,\text{TV}}$. The blue line shows the simulated instantaneous magnitude, while the red line represents the analytical envelope. (e) Normalized magnitudes of Floquet harmonics of the wave inside the intermodal band gap for the $\beta$ value selected in (d). (f)–(g) Band structures showing intermodal band gaps (highlighted in yellow) with different inclinations. (h)–(i) Corresponding simulated net energy flux $S_z$ inside the waveguide along the propagation direction for the cases in (f) and (g), respectively. In both (f) and (g), the modulation depth is $m = 0.3$ and $C_{01} = C_{02} = 0.1$~pF, while the modulation frequencies are 18~GHz for (f, h) and 24~GHz for (g, i).}
    \label{fig:unique}
\end{figure}



\subsubsection{Intermodal band gaps and directional amplification}

Next, we turn  our attention to the intermodal band gaps. Here, we choose \(d=10~\mathrm{mm}\), \(C_{01}=C_{02}=0.1~\mathrm{pF}\), \(\omega_\mathrm{m}=2\pi\times6~\mathrm{GHz}\), \(\phi=0\), and \(m_1=m_2=0.3\).
The corresponding band structure and the zoomed-in view of the intermodal band gap are shown in Figs.~\ref{fig:unique}(a), (b), and (d). 
It can be seen that such band gaps arise only at the  intersections of different mode branches.
They always occur in pairs (with real parts of the Floquet frequencies $\omega$ having opposite signs) and exhibit an exotic \textit{tilted} profile.   Here, the term \textit{tilted} refers to the inclination of $\mathrm{Re}(\omega)$ within the band gap in the $\mathrm{Re}(\omega)$--$\beta$ dispersion diagram, indicating that the real part of the complex Floquet eigenfrequency varies with the propagation constant $\beta$. Similar non-flat, tilted band gaps have been reported in Lorentzian dispersive time-varying bulk media, where they originate from the coupling between different polaritonic branches under temporal modulation~\cite{feng2024temporal,ozlu2025floquet}. In contrast, here the tilted gaps originate from intermodal coupling between guided modes in a time-modulated metasurface waveguide, rather than from volumetric Lorentzian material dispersion. Compared with the reported volumetric dispersive PTC platforms, the present guided-wave metasurface configuration provides the most straightforward and experimentally accessible route to tilted band gaps, since the required coupling is engineered through time-modulated surface reactances rather than through modulation of a bulk dispersive medium. Importantly, unlike conventional intramodal band gaps that occur at $\mathrm{Re}(\omega)/\omega_m = 0.5$, these tilted intermodal gaps are not constrained to this condition.

In Figs.~\ref{fig:unique}(c) and (e), we  show the simulated results for a wave propagating inside the tilted band gap.  
We selected the wave number of $0.39\beta_{\mathrm{m}}$, with a real part of the eigenfrequency equal to $0.25\omega_{\mathrm{m}}$ inside the intermodal band gap. Using the same simulation setup as that employed for intramodal band gap, we  obtained the evolution of the electric field. It can be seen that the simulated growth rate of the electric field again matches the theoretical values perfectly.

Due to the tilted profile of the intermodal band gap, the energy velocity of the waves inside it is nonzero. Consequently, the wave inside the intermodal band gap can be amplified in time while still exhibiting spatial propagation. To better understand this behavior, we calculate the harmonic distribution of the eigenmode within an intermodal band gap, as shown in Fig.~\ref{fig:unique}(e). In contrast to the symmetric harmonic distribution of a wave inside the intramodal band gap, Fig.~\ref{fig:unique}(e) shows a strongly asymmetric distribution. This asymmetry  indicates that the field inside the intermodal band gap is not a standing wave, in agreement with its nonzero energy velocity. This directional amplification phenomenon is in stark contrast to the intramodal band gap in conventional PTCs. 

Furthermore, we find that the direction of the net energy flux does not necessarily coincide with the group velocity. Here, in the case of intermodal gap where the Floquet eigenfrequency does not satisfy $2\omega/\omega_m \in \mathbb{Z}$, the net energy flux can be calculated as the sum of the time-averaged Poynting fluxes of all Floquet harmonics consistent with the time-domain analysis in~\cite{zurita2009reflection}. Remarkably, a nonzero net energy flux may still exist even when the group velocity is zero. In Fig.~\ref{fig:unique}(f), the real eigenfrequency inside the gap forms a straight line with positive slope, i.e., a positive group velocity. This is consistent with the asymmetric harmonic distribution in Fig.~\ref{fig:unique}(e), which shows that the eigenmode is dominated by one set of Floquet harmonics rather than by two equally weighted symmetry-related components. Thus, the relevant group velocity is determined by the Floquet branch selected by the excitation. However, the time-averaged Poynting flux is not uniform across the waveguide cross section: it is strongly negative in the central region, while weak positive contributions appear near the two metasurfaces. After integrating over the transverse direction, the resulting net energy flux is still negative. Similarly, in Fig.~\ref{fig:unique}(g), although the group velocity is nearly zero, the transversely integrated net energy flux remains nonzero and negative. This behavior arises because multiple Floquet harmonics generated by the periodic time modulation contribute differently to the local energy flow, leading to partial cancellation across the waveguide cross section but a finite negative net flux overall. More importantly, we empirically found that the degree of inclination of the intermodal band gap is determined by the group velocity difference of the corresponding branches in the static dispersion curve. This finding is similar to the conclusion in \cite{mealy2023exceptional} (Equation 11). However, \cite{mealy2023exceptional} considers periodic modulation in space, not in time. Through temporal coupled mode theory, we prove that this conclusion is also applicable to PTCs. Detailed derivation can be found in Appendix C.

\subsubsection{Temporal glide symmetry effects}

\begin{figure}[h]
    \centering
    \includegraphics[width=1\linewidth]{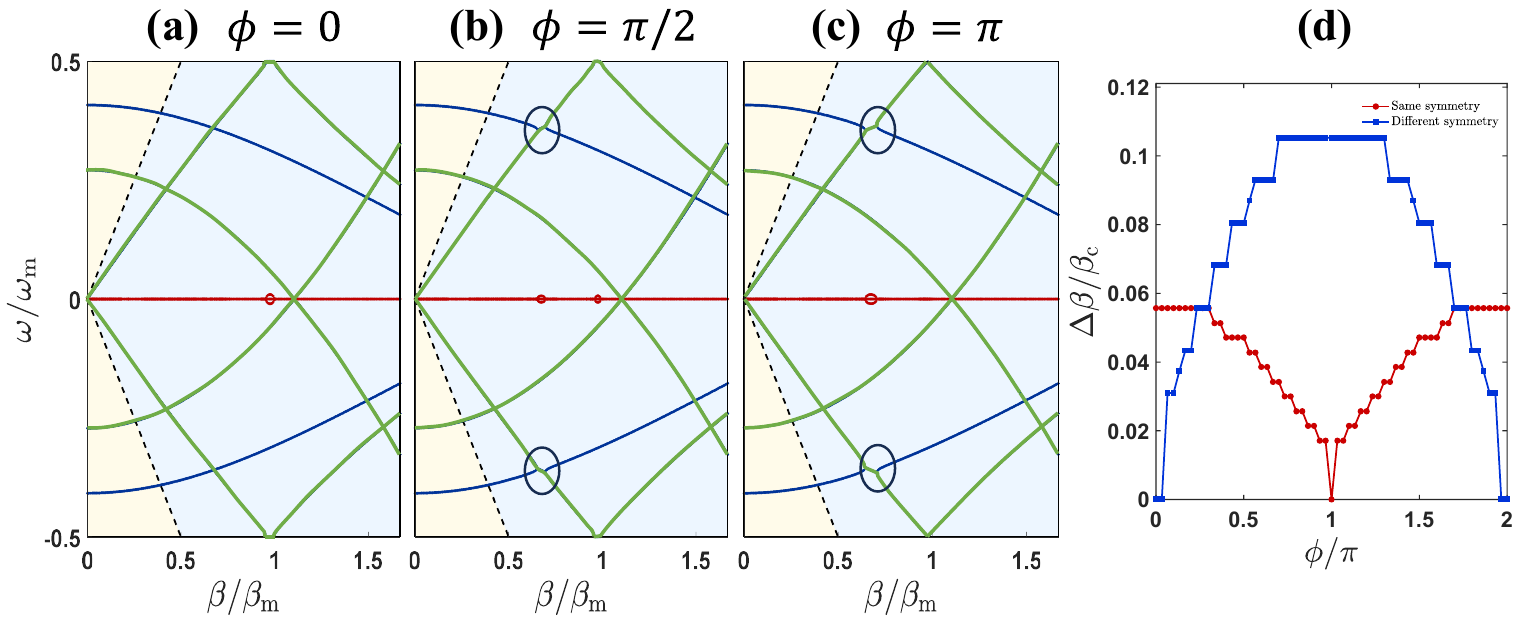}
    \caption{(a)--(c) Band structures of the PTCs for different phase shifts $\phi$ of the capacitance modulation functions of the two metasurfaces. The green lines represent even modes, while the blue lines represent odd modes. The opening and closing of representative band gaps are marked by black ellipses. (d) Relative band-gap size $\Delta\beta/\beta_{\mathrm{c}}$ as a function of $\phi$, where $\Delta\beta$ is the band-gap width and $\beta_{\mathrm{c}}$ is the wavenumber at the band-gap center. The red curve corresponds to band gaps formed by same-symmetry mode interactions, while the blue curve corresponds to band gaps formed by different-symmetry mode interactions.}
    \label{fig:8}
\end{figure}

In the above discussion, we assumed that the modulation phase difference between the two metasurfaces is  $\phi=0$. In fact, however, the phase difference has a crucial influence on the band structure. Figure~\ref{fig:8} shows the corresponding band structure for $\phi=0$, $\pi/2$, and $\pi$, respectively. The changes in the band structure are marked by black ellipses. When $\phi=0$, the intersections between different guided surface branches remain gapless, while the intersection between  the lower surface mode branch and the first-order guided volume branch produces a small band gap. At $\phi=\pi/2$, some previously closed intermodal band gaps open, whereas the intramodal band gap becomes smaller. Finally, when $\phi=\pi$, the intramodal band gap closes completely while the intermodal band gaps open maximally. 

The closure of the intramodal band gaps can be understood in terms of parametric capacitance modulation. Consider two time-varying capacitors: if one capacitor suddenly decreases its capacitance from its maximum value to its minimum at a certain moment, the voltage across it increases abruptly, and therefore the stored energy also increases accordingly. In contrast, the other capacitor behaves oppositely, suddenly increasing from its minimum value to maximum resulting in a decrease in stored energy. When  two capacitors have  equal modulation amplitudes and opposite phases, the modulation process becomes gain-loss balanced,  leading naturally to the closure of the band gap.
However, this rule does not apply to intermodal band gaps, since they are formed by the coupling of different modes and do not necessarily occur at half the modulation frequency.

More fundamentally, the present system is itself a temporally glide-symmetric structure. 
Glide symmetry is a concept usually associated with spatially periodic structures, which are invariant under a reflection followed by a translation by half a \textit{spatial} period. Such symmetry has been widely used to engineer band gaps and dispersion in electromagnetic periodic structures \cite{7302006, yuan2019glide, 8963251}, with  important applications in metasurface lenses and antennas. The temporal counterpart of this concept, namely, temporal glide symmetry, corresponds in our case to invariance under the reflection   followed by a translation by half a \textit{temporal} period. This symmetry has previously been discussed in topological systems where it was shown to protect Floquet topological phases~\cite{morimoto2017floquet}, suppressing  symmetric same-branch coupling while maximizing inter-branch coupling.
In our system, this condition is satisfied when the two sheet impedances are identical with modulation  phase difference $\phi=\pi$.  
Next, we study the band structure of such a temporal glide symmetric system.
To make this property explicit, we need to exploit the symmetry of the system. Thus, instead of the forward–backward transverse-wave basis, we employ a specific basis transformation matrix to recast Eq.~(\ref{eq.6}) into the even–odd mode basis, i.e.,
\begin{equation}
\begin{aligned}
\left[
\begin{array}{c}
    H_\mathrm{e}\\
    H_\mathrm{o}\\
\end{array}
\right]=
\begin{bmatrix}
\Bar{\Bar{I}} & \Bar{\Bar{I}} \\
\Bar{\Bar{I}} & -\Bar{\Bar{I}}
\end{bmatrix}
\left[
\begin{array}{c}
    H_+\\
    H_-\\
\end{array}
\right].
\end{aligned}
\end{equation}
Here, $H_\mathrm{e}$ and $H_\mathrm{o}$ are column vectors that represent the amplitudes of even and odd modes, respectively. If the modulation strength $m$ is much smaller than unity, then the higher order harmonics can be neglected, leaving the $-1$ and 0 order harmonics remaining. Then the eigenmatrix Eq.~(\ref{eq.15}) will now transform into the new form of:
\begin{equation}
\Bar{\Bar{K}}=\resizebox{0.91\textwidth}{!}{$
\begin{bmatrix} 
C_{0}+j\omega_{0}c_{0}\frac{j\omega_{0}\mu_{0}}{k_{c,0}}S_{0} 
&
j\omega_{0}\frac{m}{2}c_{0}\frac{j\omega_{-1}\mu_{0}}
{k_{c,-1}}S_{-1}e^{j\frac{\varphi}{2}}
&
S_{0}-j\omega_{0}c_{0}\frac{j\omega_{0}\mu_{0}}{k_{c,0}}C_{0}
&
-j\omega_{0}\frac{m}{2}c_{0}\frac{j\omega_{-1}\mu_{0}}{k_{c,-1}}C_{-1}e^{j\frac{\varphi}{2}}
\\
j\omega_{-1}\frac{m}{2}c_{0}\frac{j\omega_{0}\mu_{0}}{k_{c,0}}S_{0}e^{-j\frac{\varphi}{2}}
&
C_{-1}+j\omega_{-1}c_{0}\frac{j\omega_{-1}\mu_{0}}{k_{c,-1}}S_{-1}
&
-j\omega_{-1}\frac{m}{2}c_{0}\frac{j\omega_{0}\mu_{0}}{k_{c,0}}C_{0}e^{-j\frac{\varphi}{2}}
&
S_{-1}-j\omega_{-1}c_{0}\frac{j\omega_{-1}\mu_{0}}{k_{c,-1}}C_{-1}
\\
C_{0}+j\omega_{0}c_{0}\frac{j\omega_{0}\mu_{0}}{k_{c,0}}S_{0} 
&
j\omega_{0}\frac{m}{2}c_{0}\frac{j\omega_{-1}\mu_{0}}{k_{c,-1}}S_{-1}e^{-j\frac{\varphi}{2}} 
&
-S_{0}+j\omega_{0}c_{0}\frac{j\omega_{0}\mu_{0}}{k_{c,0}}C_{0} 
&
j\omega_{0}\frac{m}{2}c_{0}\frac{j\omega_{-1}\mu_{0}}{k_{c,-1}}C_{-1}e^{-j\frac{\varphi}{2}} 
\\
j\omega_{-1}\frac{m}{2}c_{0}\frac{j\omega_{0}\mu_{0}}{k_{c,0}}S_{0}e^{j\frac{\varphi}{2}} 
&
C_{-1}+j\omega_{-1}c_{0}\frac{j\omega_{-1}\mu_{0}}{k_{c,-1}}S_{-1} 
& 
j\omega_{-1}\frac{m}{2}c_{0}\frac{j\omega_{0}\mu_{0}}{k_{c,0}}C_{0}e^{j\frac{\varphi}{2}} 
&
-S_{-1}+j\omega_{-1}c_{0}\frac{j\omega_{-1}\mu_{0}}{k_{c,-1}}C_{-1}
\end{bmatrix},
$}
\end{equation}
where \(C_{n}=\mathrm{cos}(k_{c,n}\frac{d}{2})\), \(S_{n}=\mathrm{sin}(k_{c,n}\frac{d}{2})\), \(c_{0}\) is the average capacitance. One thing that can be seen from this matrix is that it can be decomposed into a static part and a perturbation part. The static part consists of the elements that are not associated with time modulation, i.e., those independent of $m$ or $\phi$. While the perturbation part consists of the remaining elements that arise from the weak time modulation. After a specific row operation, which is
\begin{equation}
\bar{\bar{R}}=\frac{1}{2}
\begin{bmatrix}
    1 &0 &1 &0\\
    0 &1 &0 &1\\
    1 &0 &-1 &0\\
    0 &1 &0 &-1
\end{bmatrix},
\end{equation}
we can rewrite the matrix into the following form without changing its properties:
\begin{equation}
\bar{\bar{K}}^\prime =\bar{\bar{R}} \bar{\bar{K}}=\begin{bmatrix}
    \Bar{\Bar{M}}_{\mathrm{se}}+m c_0 \cos(\phi/2)\Bar{\Bar{M}}_{\mathrm{ee}} &jm c_0 \sin(\phi/2)\Bar{\Bar{M}}_{\mathrm{eo}}\\
    jm c_0 \sin(\phi/2)\Bar{\Bar{M}}_{\mathrm{oe}} &\Bar{\Bar{M}}_{\mathrm{so}}+m c_0 \cos(\phi/2)\Bar{\Bar{M}}_{\mathrm{oo}}
\end{bmatrix}.
\end{equation} 
Here, $\bar{\bar{K}}$ and $\bar{\bar{K}}^\prime$ represent the eigenmatrix under even-odd mode basis and after a specific row operation, respectively. The $\Bar{\Bar{M}}_{\mathrm{se}}$ and $\Bar{\Bar{M}}_{\mathrm{so}}$ denote the even and odd blocks of the static matrix, while $\Bar{\Bar{M}}_{\mathrm{ee}}$, $\Bar{\Bar{M}}_{\mathrm{eo}}$, $\Bar{\Bar{M}}_{\mathrm{oe}}$ and $\Bar{\Bar{M}}_{\mathrm{oo}}$ denote the coupling matrices between even-even, even-odd, and odd-odd, respectively. Now, we can easily see that when the phase difference is zero, the off-diagonal block matrices vanish. In this condition, the even and odd modes are fully decoupled and self-coupling is maximized through $\cos(\phi/2)$. As a result, only crossings between bands of the same symmetry can open a momentum band gap. In contrast, when the phase difference is $\phi=\pi$, the self-coupling matrices $\Bar{\Bar{M}}_{\mathrm{ee}}$ and $\Bar{\Bar{M}}_{\mathrm{oo}}$ vanish, so crossings between bands of the same symmetry can no longer open a momentum band gap. While the cross-coupling matrices remain and are maximized through $\sin(\phi/2)$.

Now, we have a deeper physical insight into the time-varying metasurface waveguide. We can conclude that if the waveguide is rigorously symmetric, meaning that the two sheet impedances are identical, then the size of the momentum band gap is controlled by the modulation difference according to the following rule:
\begin{equation}
    \Delta \beta_\mathrm{same-symmetry} \propto |\cos(\phi/2)|,\quad \Delta \beta_\mathrm{different-symmetry} \propto |\sin(\phi/2)|,
\end{equation}
where $\Delta \beta_\mathrm{same-symmetry}$ and $\Delta \beta_\mathrm{different-symmetry}$ represent the size of the momentum band gap formed by the same symmetric and different symmetric bands, respectively, as Fig.~\ref{fig:8}(d) shows. The phase difference provides a powerful degree of freedom to tune the band structure beyond the modulation strength. Changing the modulation strength can indeed enlarge or shrink the momentum band gaps, but the intramodal band gap and the intermodal band gap change simultaneously. In contrast, varying the phase difference can tune the intramodal and intermodal band gaps in an opposite manner. Physically speaking, the phase difference provides a way to mimic—and continuously tune—the effective spatial symmetry of the waveguide or any coupled system in the time domain. Thus, from a practical implementation point of view, it also offers a convenient route to realize more complex functionalities, such as a momentum-selective amplifier.

\section{CONCLUSION}

In this work, we have investigated a multimode waveguide PTC formed by two impenetrable metasurfaces with time-varying surface capacitances. We first derived the dispersion relation of the corresponding stationary waveguide and then used it as the basis for calculating the band structure of the time-modulated system. The results show that this artificial multimode electromagnetic platform supports not only conventional intramodal momentum band gaps, but also tilted intermodal momentum band gaps arising from coupling between different guided-mode branches.
Compared with previously studied time-varying dispersive bulk media, the proposed metasurface waveguide offers greater design flexibility and a more experimentally accessible platform for wave manipulation, because multiple guided modes of different physical character can be engineered within the same structure. By tuning the modulation frequency, modulation depth, and modulation phase difference between the two metasurfaces, one can systematically control the opening and closing of both conventional and tilted band gaps, as well as their bandwidths and tilt angles.
A distinctive feature of the tilted intermodal momentum band gaps is that the amplified waves inside them are not standing waves. Instead, they can possess a nonzero energy flux along the waveguide, providing a route toward active guided-wave energy routing and amplification. Moreover, by exploiting the temporal glide symmetry of the system, we established a selection rule that governs the opening or closing of gaps associated with same-symmetry and different-symmetry mode interactions.
Overall, our findings extend PTCs from single- or dual-mode platforms to a genuinely multimode regime, opening new opportunities for exploring emergent phenomena in time-varying electromagnetics.

Although this work focuses on PTCs based on impenetrable metasurface waveguides, this analysis can be readily extended to penetrable metasurface waveguides, which, in the stationary regime, have been shown to support surface modes together with leaky free-space radiation channels rather than the fully confined guided volume modes considered here~\cite{ma2019parallel,hosseini2023study}. Such penetrable configurations may also provide an alternative route to metasurface-based PTCs with qualitatively different mode coupling, radiation, and amplification characteristics. 
% ...

Regarding experimental feasibility at microwave frequencies, metasurfaces can be readily fabricated using standard printed-circuit-board (PCB) technology, while temporal and even spatiotemporal modulations can be implemented with varactor diodes~\cite{wu2020space,wang2023metasurface,ye2025realization,li2025emulating}. At terahertz or even higher frequency regimes, the proposed impenetrable metasurface waveguide can be realized by leveraging ultrafast carrier dynamics in small band-gap semiconductors or transparent conductive oxides (TCOs). Specifically, recent experimental breakthroughs in single-layer plasmonic time crystal using InSb-based cavities have demonstrated that field-induced modulation of effective mass can yield substantial and sub-optical-cycle variations in capacitance~\cite{guo2025plasmonic}. By arranging two such metasurfaces in a face-to-face configuration, the predicted tilted momentum band gap and directional amplification could be experimentally verified at terahertz or infrared frequencies. Such a platform, synergizing the robustness of metasurface design with the richness of Floquet physics, paves the way for a new generation of on-chip active components and ultrafast time-domain photonic devices.

\bibliography{reference}

\section{Appendix A}
\begin{figure}[t]
    \centering
    \includegraphics[width=1\linewidth]{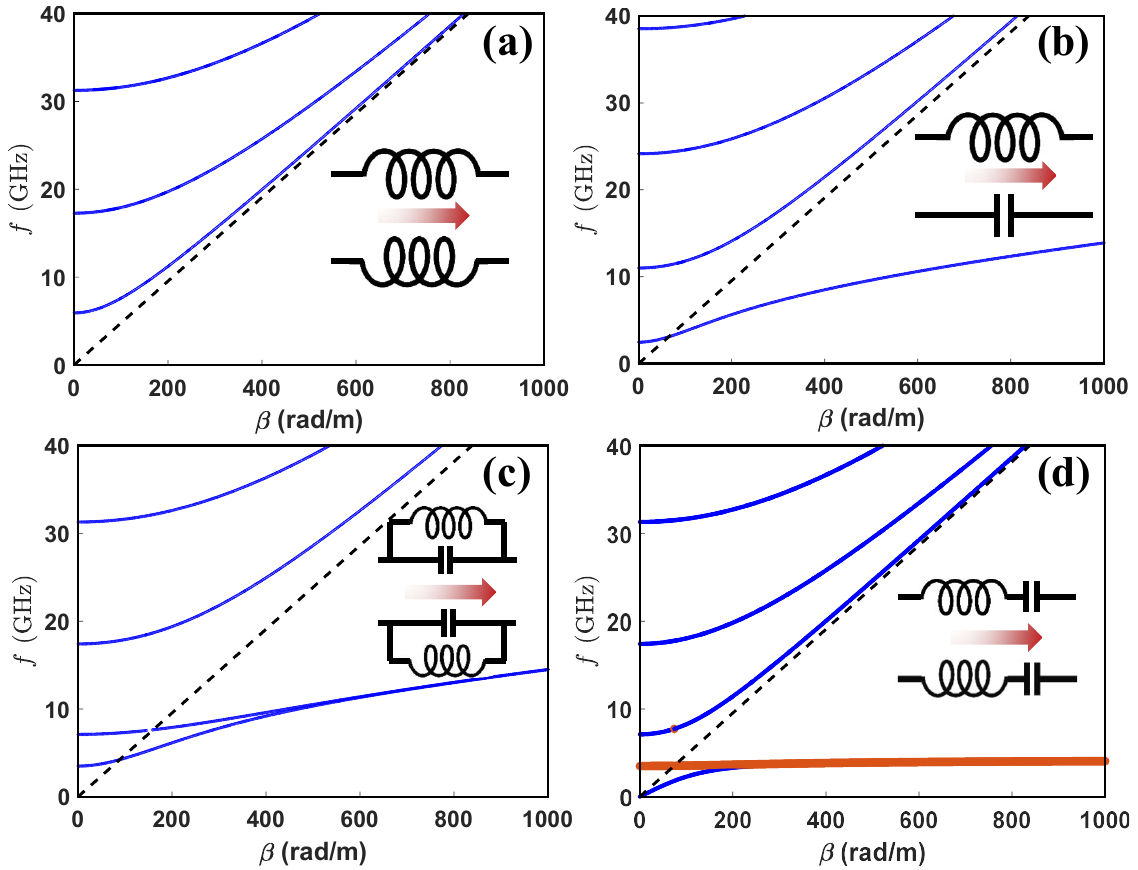}
\caption{Dispersion curves of the waveguide formed by two impenetrable time-invariant metasurfaces with different types of sheet impedance. (a)–(b) Non-resonant cases: (a) two inductive sheets, $L_1=L_2=14.2~\mathrm{nH}$; (b) one inductive and one capacitive sheet, $C_1=0.1~\mathrm{pF}$ and $L_2=14.2~\mathrm{nH}$. (c)–(d) Resonant cases: (c) two parallel LC sheets and (d) two series LC sheets, with $C_1=C_2=0.1~\mathrm{pF}$ and $L_1=L_2=14.2~\mathrm{nH}$. In all cases, $d=10~\mathrm{mm}$.}
    \label{fig:3}
\end{figure}

In the main text, we just consider waveguide formed by two capacitive metasurfaces. Without loss of generality, we further consider two metasurfaces with different types of impedance beyond purely capacitive reactance, including two inductive sheets, one inductive and one capacitive sheet, and two LC resonant sheets in this section. The theory for TE eigenmodes developed in the main text is general, and the corresponding dispersion relations can be obtained simply by replacing  $Z_1$ and $Z_2$ in Eq.~(\ref{eq.4}) with the desired sheet impedances. The resulting dispersion curves are shown in Fig.~\ref{fig:3}. As seen from Fig.~\ref{fig:3}(a), when both metasurfaces are inductive, no guided surface modes exist, since an inductive metasurface does not support TE surface modes. Nevertheless, the guided volume modes survive. Figure~\ref{fig:3}(b) shows that when the waveguide is formed by one inductive and one capacitive metasurface, only one surface mode branch can survive which corresponds to  the odd mode due to the opposite-signs of the two reactances. These two waveguide configurations are non-resonant. 

By contrast, Figs.~\ref{fig:3}(c) and (d) correspond to  resonant configurations. In Fig.~\ref{fig:3}(c), the two metasurfaces can be modeled as parallel LC circuits, whereas in Fig.~\ref{fig:3}(d) they are modeled as series LC circuits. Notably, when the two series LC metasurfaces are identical, an extremely flat band over the entire $\beta$-space emerges (indicated by orange line). This flat band corresponds to  the odd mode. This can be mathematically proven by only considering the odd mode in Eqs.~(\ref{eq.1}) and (\ref{eq.2}) and obtaining the corresponding dispersion relation which is nearly $\beta$-independent. Such an all-$\beta$ flat band indicates interesting wave physics and advanced functionalities: it implies the group velocity of surface modes is approximately zero, allowing the appearance of a macroscopically degenerate set of odd guided surface modes with a large density of states and strong confinement of energy between the two metasurfaces. 
More importantly, the odd mode is radiatively dark (BIC-like), so that finite structures inherit ultra-high-Q, angle-robust resonances and strong light–matter interaction which has promising applications for nonlinear and  switching applications. 

\section{Appendix B}
This section shows the detailed matrix used in the main text. Equation~(\ref{eq.11}a) can be rewritten as:
\begin{equation}
\begin{aligned}
&\begin{bmatrix}
    &e^{-j \sqrt{\omega_N^2 \epsilon_0 \mu_0 -\beta^2}d} &0 &\cdots &0\\
    &0 &e^{-j \sqrt{\omega_{N-1}^2 \epsilon_0 \mu_0 -\beta^2}d} &\cdots &\vdots\\
    &\vdots &\vdots &\ddots &\vdots\\
    &0 &0 &\cdots &e^{-j \sqrt{\omega_{-N}^2 \epsilon_0 \mu_0 -\beta^2}d}
\end{bmatrix}
\left[
\begin{array}{c}
    H_{+,N}\\
    H_{+,N-1}\\
    \vdots\\
    H_{+,-N}
\end{array}
\right]\\
&+\begin{bmatrix}
    &e^{j \sqrt{\omega_N^2 \epsilon_0 \mu_0 -\beta^2}d} &0 &\cdots &0\\
    &0 &e^{j \sqrt{\omega_{N-1}^2 \epsilon_0 \mu_0 -\beta^2}d} &\cdots &\vdots\\
    &\vdots &\vdots &\ddots &\vdots\\
    &0 &0 &\cdots &e^{j \sqrt{\omega_{-N}^2 \epsilon_0 \mu_0 -\beta^2}d}
\end{bmatrix}
\left[
\begin{array}{c}
    H_{-,N}\\
    H_{-,N-1}\\
    \vdots\\
    H_{-,-N}
\end{array}
\right]\\
&=B
\begin{bmatrix}
    &\frac{\omega_{N}c_{1(0)}e^{-j k_{c(N)}d}}{\sqrt{\omega_N^2 \epsilon_0 \mu_0 -\beta^2}} &\frac{\omega_{N-1}c_{1(1)}e^{-j k_{c(N-1)}d}}{\sqrt{\omega_{N-1}^2 \epsilon_0 \mu_0 -\beta^2}} &\cdots &\frac{\omega_{-N}c_{1(2N)}e^{-j k_{c(-N)}d}}{\sqrt{\omega_{-N}^2 \epsilon_0 \mu_0 -\beta^2}}\\
    &\frac{\omega_{N}c_{1(-1)}e^{-j k_{c(N)}d}}{\sqrt{\omega_{N}^2 \epsilon_0 \mu_0 -\beta^2}} &\frac{\omega_{N-1}c_{1(0)}e^{-j k_{c(N-1)}d}}{\sqrt{\omega_{N-1}^2 \epsilon_0 \mu_0 -\beta^2}} &\cdots &\frac{\omega_{-N}c_{1(2N-1)}e^{-j k_{c(-N)}d}}{\sqrt{\omega_{-N}^2 \epsilon_0 \mu_0 -\beta^2}}\\
    &\vdots &\vdots &\ddots &\vdots\\
    &\frac{\omega_{N}c_{1(-2N)}e^{-j k_{c(N)}d}}{\sqrt{\omega_{N}^2 \epsilon_0 \mu_0 -\beta^2}} &\frac{\omega_{N-1}c_{1(-2N+1)}e^{-j k_{c(N-1)}d}}{\sqrt{\omega_{N-1}^2 \epsilon_0 \mu_0 -\beta^2}} &\cdots &\frac{\omega_{-N}c_{1(0)}e^{-j k_{c(-N)}d}}{\sqrt{\omega_{-N}^2 \epsilon_0 \mu_0 -\beta^2}}
\end{bmatrix}
\left[
\begin{array}{c}
    H_{+,N}\\
    H_{+,N-1}\\
    \vdots\\
    H_{+,-N}
\end{array}
\right]\\
&+B
\begin{bmatrix}
    &\frac{\omega_{N}c_{1(0)}e^{j k_{c(N)}d}}{\sqrt{\omega_N^2 \epsilon_0 \mu_0 -\beta^2}} &\frac{\omega_{N-1}c_{1(1)}e^{j k_{c(N-1)}d}}{\sqrt{\omega_{N-1}^2 \epsilon_0 \mu_0 -\beta^2}} &\cdots &\frac{\omega_{-N}c_{1(2N)}e^{j k_{c(-N)}d}}{\sqrt{\omega_{-N}^2 \epsilon_0 \mu_0 -\beta^2}}\\
    &\frac{\omega_{N}c_{1(-1)}e^{j k_{c(N)}d}}{\sqrt{\omega_{N}^2 \epsilon_0 \mu_0 -\beta^2}} &\frac{\omega_{N-1}c_{1(0)}e^{j k_{c(N-1)}d}}{\sqrt{\omega_{N-1}^2 \epsilon_0 \mu_0 -\beta^2}} &\cdots &\frac{\omega_{-N}c_{1(2N-1)}e^{j k_{c(-N)}d}}{\sqrt{\omega_{-N}^2 \epsilon_0 \mu_0 -\beta^2}}\\
    &\vdots &\vdots &\ddots &\vdots\\
    &\frac{\omega_{N}c_{1(-2N)}e^{j k_{c(N)}d}}{\sqrt{\omega_{N}^2 \epsilon_0 \mu_0 -\beta^2}} &\frac{\omega_{N-1}c_{1(-2N+1)}e^{j k_{c(N-1)}d}}{\sqrt{\omega_{N-1}^2 \epsilon_0 \mu_0 -\beta^2}} &\cdots &\frac{\omega_{-N}c_{1(0)}e^{j k_{c(-N)}d}}{\sqrt{\omega_{-N}^2 \epsilon_0 \mu_0 -\beta^2}}
\end{bmatrix}
\left[
\begin{array}{c}
    -H_{-,N}\\
    -H_{-,N-1}\\
    \vdots\\
    -H_{-,-N}
\end{array}
\right].
\end{aligned}
\end{equation}

And Eq.~(\ref{eq.11}b) can also be rewritten as:

\begin{equation}
\begin{aligned}
&\left[
\begin{array}{c}
    H_{+,N}+H_{-,N}\\
    H_{+,N-1}+H_{-,N-1}\\
    \vdots\\
    H_{+,-N}+H_{-,-N}
\end{array}
\right]
=
\begin{bmatrix}
    &j\omega_N \mu_0 &0 &\cdots &0\\
    &0 &j\omega_{N-1} \mu_0 &\cdots &\vdots\\
    &\vdots &\vdots &\ddots &\vdots\\
    &0 &0 &\cdots &j\omega_{-N} \mu_0
\end{bmatrix}\\
&\begin{bmatrix}
    &\frac{\omega_{N}c_{2(0)}}{\sqrt{\omega_N^2 \epsilon_0 \mu_0 -\beta^2}} &\frac{\omega_{N-1}c_{2(1)}}{\sqrt{\omega_{N-1}^2 \epsilon_0 \mu_0 -\beta^2}} &\cdots &\frac{\omega_{-N}c_{2(2N)}}{\sqrt{\omega_{-N}^2 \epsilon_0 \mu_0 -\beta^2}}\\
    &\frac{\omega_{N}c_{2(-1)}}{\sqrt{\omega_{N}^2 \epsilon_0 \mu_0 -\beta^2}} &\frac{\omega_{N-1}c_{2(0)}}{\sqrt{\omega_{N-1}^2 \epsilon_0 \mu_0 -\beta^2}} &\cdots &\frac{\omega_{-N}c_{2(2N-1)}}{\sqrt{\omega_{-N}^2 \epsilon_0 \mu_0 -\beta^2}}\\
    &\vdots &\vdots &\ddots &\vdots\\
    &\frac{\omega_{N}c_{2(-2N)}}{\sqrt{\omega_{N}^2 \epsilon_0 \mu_0 -\beta^2}} &\frac{\omega_{N-1}c_{2(-2N+1)}}{\sqrt{\omega_{N-1}^2 \epsilon_0 \mu_0 -\beta^2}} &\cdots &\frac{\omega_{-N}c_{2(0)}}{\sqrt{\omega_{-N}^2 \epsilon_0 \mu_0 -\beta^2}}
\end{bmatrix}
\left[
\begin{array}{c}
    -H_{+,N}+H_{-,N}\\
    -H_{+,N-1}+H_{-,N-1}\\
    \vdots\\
    -H_{+,-N}+H_{-,-N}
\end{array}
\right].
\end{aligned}
\end{equation}

\section{Appendix C}
To explain why the band gap is tilted and which parameters are associated with the tilt angle, we then analytically calculate the tilt angle within the framework of temporal coupled mode theory.
We begin by considering a configuration where the dispersive band diagram supports two bands, both exhibiting positive group velocity. Under temporal modulation, one of these bands evolves into a symmetry-related branch with negative frequency. Owing to Floquet periodicity, this branch can be folded by an integer multiple of the modulation frequency, so that it overlaps with the other band and becomes coupled through the time-periodic modulation. The resulting hybridization opens a momentum band gap. In what follows, we investigate the slope of the real part of the complex Floquet eigenfrequency inside the tilted band gap and establish its relation to the group velocities of the two underlying static bands.
\\
\indent To address the issue, we first denote the dispersion relations of the two static bands as $\omega_1(\beta)$ and $\omega_{2}(\beta)$.  Once the modulation is applied and the energy is injected into the system, $\omega_{2}(\beta)$ undergoes symmetry and shifting transformation, after time reversal, to $n\Omega-\omega_{2}(\beta)$. Here $n$ is the Floquet harmonic order. The coupling between the two bands can be expressed as:
\begin{equation}
i\,\frac{d}{dt}
\begin{pmatrix}
a(t) \\
b(t)
\end{pmatrix}
=
\underbrace{
\begin{pmatrix}
\omega_{1}(\beta) & g_{1,2} \\
g_{2,1} & n\Omega-\omega_{2}(\beta)
\end{pmatrix}
}_{H_{\mathrm{eff}}(\beta)}
\begin{pmatrix}
a(t) \\
b(t)
\end{pmatrix},
\label{eq:TCMT function}
\end{equation}
Here $a(t)$ and $b(t)$ are the amplitudes of the Floquet channels. In a spatially homogeneous medium, they reduce to time-dependent factors. The mode-degeneracy condition is given by
\begin{equation}
\omega_1(\beta)\approx n\Omega-\omega_{2}(\beta)
\label{eq:degeneracy condition}
\end{equation}
The effective coupling strength between the two modes is denoted by $g$
(with units of $\mathrm{rad/s}$).
The corresponding $2\times2$ eigenvalue problem can be written as
\begin{equation}
\det\!\left(
\begin{pmatrix}
\omega_{1}(\beta) & g_{1,2} \\
g_{2,1} & n\Omega-\omega_{2}(\beta)
\end{pmatrix}
- \omega I
\right) = 0 .
\end{equation}
Defining
\begin{equation}
\Delta(\beta) = \omega_{1}(\beta)-\omega_{2}(\beta) + n\Omega, 
\qquad
\Sigma(\beta) = \omega_{1}(\beta) +\omega_{2}(\beta)- n\Omega,
\end{equation}
the eigenfrequencies are given by
\begin{equation}
\omega_{\pm}(\beta)
= \frac{\Delta(\beta)}{2}
\pm
\sqrt{
\frac{\Sigma(\beta)^{2}}{4}
+ g_{1,2}g_{2,1}
}.
\label{eq:eigenfrequencie solution}
\end{equation}

For a real-valued modulation, the coefficients of the time-domain equations are real. Consequently, complex conjugation maps any solution to another valid solution, and the associated Floquet spectrum is closed. In the model we are considering, the two interacting branches are chosen to be related by this time-symmetry operation and brought into degeneracy after the appropriate frequency shifting. This symmetry constraint enforces the Bogoliubov structure of the reduced two-mode model, leading to

\begin{equation}
g_{21}=-g_{12}^{*}
\quad\Rightarrow\quad
g_{12}g_{21}=-|g_{12}|^{2}.
\label{eq:g_relation}
\end{equation}
Therefore, for purely real temporal modulation, $g_{1,2}g_{2,1}<0$, Eq.\ref{eq:eigenfrequencie solution} can be written:
\begin{equation}
\omega_{\pm}(\beta)
= \frac{\Delta(\beta)}{2}
\pm
\sqrt{
\frac{\Sigma(\beta)^{2}}{4}
-g^2
}.
\label{eq:eigenfrequencie solution new}
\end{equation}
The condition for the emergence of a temporal band gap is
\begin{equation}
|\Sigma(\beta)| < 2g .
\end{equation}

Within this interval, the square-root term becomes purely imaginary, a momentum band gap opens in the vicinity of the degeneracy point.
\begin{equation}
\sqrt{\frac{\Sigma(\beta)^{2}}{4} - g^{2}}
= i \sqrt{ g^{2} - \frac{\Sigma(\beta)^{2}}{4} }
\equiv i\,\Gamma(\beta),
\qquad
\Gamma(\beta)\in{R}_{+}.
\end{equation}
As a result, the eigenfrequencies can be written as
\begin{equation}
\omega_{\pm}(\beta)
= \frac{\Delta(\beta)}{2}
\pm i\,\Gamma(\beta)
\end{equation}
which immediately implies
\begin{equation}
\mathrm{Re}\,\omega_{\pm}(\beta)
= \frac{\omega_{1}(\beta) - \omega_{2}(\beta)+n\Omega}{2}.
\label{eq:real part fun}
\end{equation}

Taking the derivative of Eq.\ref{eq:real part fun} with respect to $\beta$, we obtain
\begin{equation}
\frac{d}{d \beta}\,\mathrm{Re}\,\omega_{\pm}(\beta)
= \frac{1}{2}
\left(
\frac{d\omega_{1}}{d\beta}
-
\frac{d\omega_{2}}{d\beta}
\right)
= \frac{v_{g1}(\beta) - v_{g2}(\beta)}{2},
\end{equation}
where $v_{g1}(\beta)$ and $v_{g2}(\beta)$ denote the group velocities of the two uncoupled modes. 
\\
Accordingly, we define the group velocity within the band gap as
\begin{equation}
V(\beta) = \frac{d}{d\beta}\,\mathrm{Re}\left\{\omega(\beta)\right\}.
\end{equation}
It then follows rigorously that, within the temporal band gap,
\begin{equation}
V(\beta) = \frac{v_{g1}(\beta) - v_{g2}(\beta)}{2}.
\end{equation}
In summary, in a purely real time-modulated system, the slope of the real part of the complex Floquet eigenfrequency inside the momentum band gap is determined by the difference between the group velocities of the two uncoupled static bands.

\end{document}